# Correcting thermal-emission-induced detector saturation in infrared reflection or transmission spectroscopy

Chunhui Yao[1], Hongyan Mei[1], Yuzhe Xiao[1], Alireza Shahsafi[1], William Derdeyn[2], Jonathan L. King[1], Chenghao Wan[1,3], Raluca O. Scarlat[2,4], Mark H. Anderson[2,5], and Mikhail A. Kats[1,3,6]

[1] *Department of Electrical and Computer Engineering, University of Wisconsin-Madison, Madison, Wisconsin 53706, USA*

[2] *Department of Engineering Physics, University of Wisconsin-Madison, Madison, Wisconsin 53706, USA*

[3] *Department of Materials Science & Engineering, University of Wisconsin-Madison, Madison, Wisconsin 53706, USA*

[4] *Department of Nuclear Engineering, University of California Berkeley, Berkeley, California 94720, USA*

[5] *Department of Mechanical Engineering, University of Wisconsin-Madison, Madison, Wisconsin 53706, USA*

[6] *Department of Physics, University of Wisconsin-Madison, Madison, Wisconsin 53706, USA*

**Abstract**

We found that temperature-dependent infrared spectroscopy measurements (i.e., reflectance or transmittance) using a Fourier-transform spectrometer can have substantial errors, especially for elevated sample temperatures and collection using an objective lens (e.g., using an infrared microscope). These errors arise as a result of partial detector saturation due to thermal emission from the measured sample reaching the detector, resulting in nonphysical apparent reduction of reflectance or transmittance with increasing sample temperature. Here, we demonstrate that these temperature-dependent errors can be corrected by implementing several levels of optical attenuation that enable "convergence testing" of the measured reflectance or transmittance as the thermal-emission signal is reduced, or by applying correction factors that can be inferred by looking at the spectral regions where the sample is not expected to have a substantial temperature dependence.

**Introduction**

Recently, new optical devices with infrared temperature-dependent properties, e.g., those based on phase-transition materials, have been used for applications that include thermal imaging [1], [2], sensing [3], [4], and passive radiative cooling [5], [6]. The infrared optical properties of such structures are typically characterized using Fourier-transform spectrometers (FTSs, a.k.a. FTIRs), or occasionally using grating-based spectrometers; in this paper, we focus on FTSs. These instruments are usually equipped with infrared detectors that have at least some nonlinearity [7]–[10], potentially resulting in experimental challenges. When the incident photon flux exceeds a certain threshold, the voltage on the detector may no longer be linearly proportional to the photon flux [11]. In FTSs, such nonlinearity distorts the interferogram [12], especially in the center-burst region where the largest variations of photon flux occur as the optical path difference passes through zero [13], [14]. But for FTS measurements of high-temperature samples, the nonlinearity may affect the entire interferogram because the detector receives not only the light from the optical source (typically a Globar) reflected or transmitted by the sample, but also thermal radiation directly emitted by the sample. This effect is particularly pronounced when the reflected or transmitted light is collected using an objective lens (e.g., in measurements using an infrared microscope), because the objective can collect thermal radiation emitted by the sample over a large solid angle in the far field.

Typical ways to correct detector nonlinearity in FTSs include both hardware [9], [15], [16] and postprocessing [12], [14], [17] approaches. Today, many commercial FTS systems provide built-in nonlinearity correction; e.g., the FTS in our laboratory (Bruker VERTEX 70) offers a software solution based on an empirical nonlinearity model whose coefficients are iteratively refined to minimize the spectral artifacts that appear below the detectable frequency (sometime called "out-of-band signal") and are understood to be caused by nonlinearity [18]. However, most of these correction approaches are designed for cases where the nonlinearity is primarily attributed to light emitted by the spectrometer source (when the thermally emitted light from the sample itself can be neglected), but cannot be directly applied to reflectance or transmittance measurements of high-temperature samples when their thermally emitted power is comparable to or even exceeds the power from the external light source.

In this work, we performed temperature-dependent spectral reflectance measurements using a commercial FTS connected to an infrared microscope and observed that the measured value of reflectance nonphysically decreases at higher sample temperatures. We found that the strong thermal radiation emitted by the hot sample partially saturates the detector, reducing its responsivity and thus resulting in the nonphysical results—even if the thermal-radiation bypasses the FTS interferometer and is not directly measured. To extract quantitatively correct spectroscopic data, we propose a "convergence testing" approach, based on several levels of attenuation that limit the thermal radiation from the sample that reaches the detector, and a data-processing approach, applying correction factors based on data from spectral regions where the sample is not expected to have substantial temperature dependence. Though we only provide experimental examples of reflectance measurements, these correction methods similarly apply to temperature-dependent transmittance measurements.

**Experiments**

**Figure 1(a)** shows the FTS setup we used for temperature-dependent reflectance measurements. The sample is placed on a temperature-controlled stage, so that its reflectance can be measured as a function of temperature. An infrared microscope is positioned after the interferometer, such that the sample is illuminated using a reflective objective (here, NA = 0.4, i.e., incident angles of about $\pm 24°$), with the reflected light collected using the same objective, and then sent to a liquid-nitrogen-cooled mercury-cadmium-telluride (MCT) detector. The light incident on the detector can be broken down into two components: (1) light from the FTS source, which passes through the interferometer and is reflected by the sample, and (2) thermal emission from the sample which bypasses the interferometer, and thus is not modulated [**Fig. 1(a)**]. We refer to the reflected light as the AC contribution, and the thermal emission from the sample as the DC contribution, because the AC coupling of the detector amplifier ensures that the thermal emission will not be directly measured. Nevertheless, this DC contribution still hits the detector, "invisibly" contributing to detector saturation.

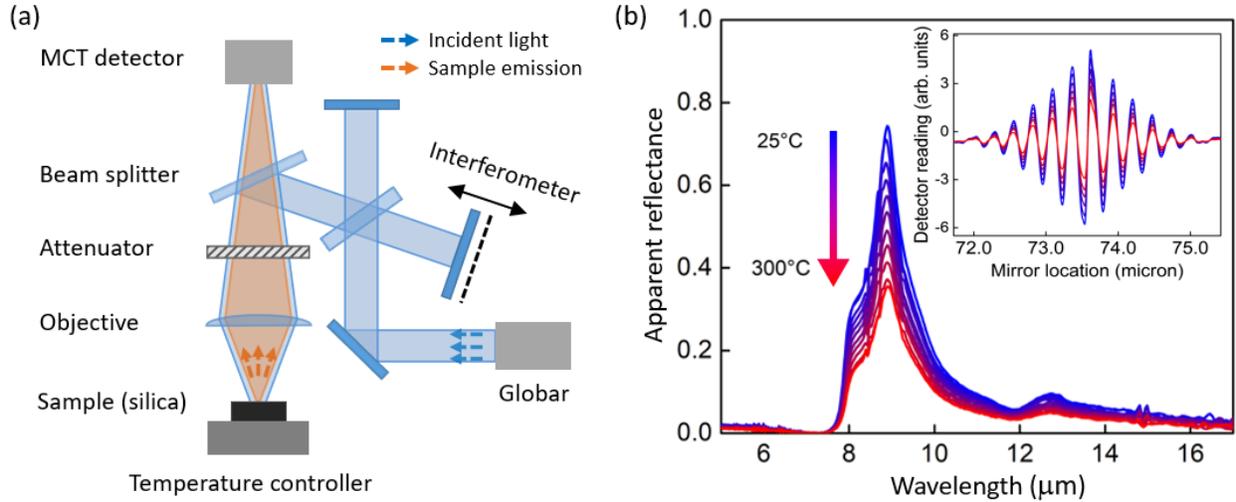

**FIG. 1. Nonlinearity in temperature-dependent reflectance measurements. (a)** The optical path of our Fourier-transform spectrometer (FTS). An infrared microscope is placed after the interferometer with a reflective objective that illuminates the sample and then collects the reflected light and sends it to the MCT detector. The photons that hit the detector mainly come from two components: (1) the light from the Globar, which passes through the interferometer and is therefore modulated, and (2) the thermal emission from the sample, which bypasses the interferometer and is therefore not modulated. **(b)** Measured apparent reflectance of a flat fused-silica wafer from 25 to 300 °C, in intervals of 25 °C, using a gold mirror (at room temperature) as a reference. We observe a (mostly) nonphysical decrease in reflectance as a function of temperature over the entire wavelength range. Inset: the center-burst regions of selected interferogram traces corresponding to the spectra in the main panel.

In **Fig. 1(b)**, we plotted the measured temperature-dependent reflectance spectra of a polished fused-silica wafer (about 0.5 millimeter thick) from 25 to 300 °C, normalized to a gold mirror. The measured reflectance is dramatically reduced for increasing temperature (more than 50% reduction from 25 to 300 °C), which is nonphysical for silica [19], and points to the presence of detector nonlinearity. The fused-silica measurement is somewhat complicated by the fact that a reflectance reduction with temperature is actually expected at frequencies near the vibrational resonance of this material [19], [20], but the expected reduction is much smaller than what is observed in **Fig. 1(b)**. Note that this anomaly includes only a reduction in the amplitude of the interferogram, but no interferogram phase shift (inset of **Fig. 1(b)**), such as the one we previously observed in certain thermal-emission measurements near room temperature [21]. We also note that it has been reported that a portion of light thermally emitted by the sample may enter the interferometer and be reflected back to the sample, adding extra distortions to the interferogram [22], [23]; we believe this path to be at most a minor effect in our setup, in particular because the reduction in reflectance in **Fig. 1(b)** appears to be wavelength-independent. To the best of our knowledge, though the nonlinear effects in FTS measurements have been widely reported, the wavelength-independent reduction in the measured interferogram for temperature-dependent microscope-based infrared reflectance and transmittance measurements has not been appropriately addressed.

**Analysis of the nonlinear response**

In MCT detectors and their associated electronics, there can be a nonlinear relationship between the recorded voltage signal and the incident photon intensity [9]. To understand the effect of the detector nonlinearity on temperature-dependent measurements, we first summarize how an interferogram is generated in an FTS: the motion of a mirror in one arm of a Michelson interferometer modifies the interference condition at the detector, resulting in a variation of incident photon flux (we refer to this mirror-position-dependent photon flux as the AC contribution). In the measurement described in **Fig. 1(a)**, photon flux on the detector also includes a DC contribution from thermal emission from the sample that bypasses the interferometer and is therefore mirror-position-independent. For the same sample at different temperatures, the AC contribution is expected to be approximately the same, since the optical properties of most materials have a weak dependence on temperature. However, the DC contribution can change substantially due to the temperature dependence of thermal emission [**Fig. 2(a)**]. **Figure 2(b, c)** schematically shows the mapping from photon flux to voltage and how the nonlinear detector responsivity affects these conversion processes. The physical position-dependent voltage (before being processed by the detector amplifier) consists of an AC part, $V^{AC}(x,T)$, and a DC part, $V^{DC}(T)$ [12], where $x$ represents the location of the moving mirror in the interferometer and $T$ is the temperature of sample. In practice, the AC-coupled detector amplifier filters out the DC part $V^{DC}(T)$, so only the AC part $V^{AC}(x,T)$ is recorded as the measured interferogram $I_m(x,T)$. However, the DC contribution of the incident photon, though typically not recorded, can nevertheless have an impact on the measurement due to detector nonlinearity. Since the detector responsivity curve is expected to be concave down with increasing photon flux (partial saturation) [**Fig. 2(b)**], the large DC contribution from a high-temperature sample can lead to a suppression in the amplitude of its recorded AC contribution [**Fig. 2(c)**]. Therefore, we introduce a sample-dependent distortion factor $d(T)$ to describe the reduction in amplitude, so that the measured interferogram can be expressed as:

$$I_m(x,T) = d(T)V^{AC}(x,T). \tag{1}$$

We expect that $d(T) = 1$ when the sample is at room temperature, but gradually decreases with increasing sample temperature, which explains the wavelength-independent decrease in our measured spectra. Note that the AC contribution can result in additional mirror-position-dependent detector saturation, especially in the center-burst region [13], [14], but this effect does not influence the mirror-position-independent amplitude suppression in the recorded interferograms induced by the DC contribution and is also not noticeable in our experiments, therefore, is not the focus of this manuscript.

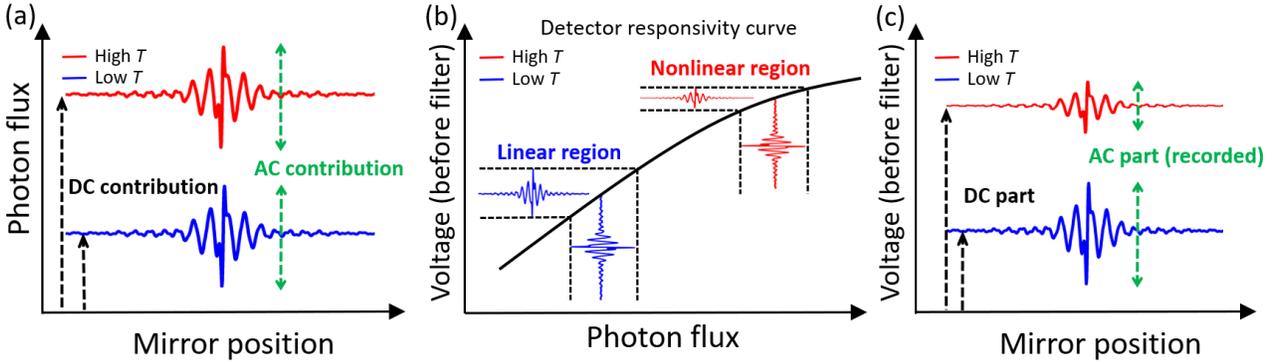

**FIG. 2. The effect of detector nonlinearity on high-temperature spectroscopy measurements**. **(a)** For high-temperature measurements in our FTS, the photon flux on the detector includes (1) the AC contribution which is mirror-position-dependent and (2) the DC contribution from the thermal emission, which is temperature-dependent. Therefore, when a sample is heated to a higher temperature, the DC contribution increases while the AC contribution remains basically unchanged. **(b)** Beyond a certain threshold, the slope of the responsivity curve (voltage vs. photon flux) decreases with increasing incident photon flux. **(c)** The voltage (before the high-pass filter) also includes the AC and DC parts, while only the AC part is recorded as the measured interferogram. Due to the decreasing slope of the responsivity curve, the unrecorded DC contribution from a high-temperature sample can cause a reduction in the amplitude of its recorded AC contribution.

**Nonlinearity correction methods and verification**

This section will introduce two nonlinearity correction methods: one based on "convergence testing" using increasing levels of attenuation in the FTS, and a second based on data processing, where we use prior knowledge about the material to correct the temperature-dependent measurement data. The two methods can be used independently or in tandem.

First, since the nonlinearity is attributed to the overlarge photon flux that strikes to the detector, the most intuitive way to minimize the nonlinearity is to introduce attenuation into the beam path [23], increasing the attenuation level until the detector response is approximately linear; we refer to this approach as **Method 1**. In our setup, we use an aluminum foil with evenly spaced holes (each several millimeters in diameter) as a mesh-type attenuator. Note that it is rather common to implement attenuations in FTS-based measuring systems to limit the intensity of incident beam from the light source. In most cases, the attenuator or optical aperture is placed between the source and the interferometer [24], [25], [26]. However, this attenuator placement will not work in our case because it is the thermal emission from the sample that bypasses the interferometer that results in the nonlinearity. Therefore, our attenuator is positioned right above the objective (**Fig. 1(a)**), such that it reduces both the reflected light (not necessarily desired) and the thermal radiation from the sample.

**Figure 3(a-c)** shows the measured apparent reflectance of fused silica from 25 to 300 °C using different levels of attenuation (no attenuation, attenuation to about 25% of the original photon flux and attenuation to about 15%, respectively) to perform a "convergence test". As the attenuation increases, the nonphysical reduction in the apparent reflectance begins to vanish. When the intensity is attenuated to about 15% in our measurement, the measured reflectance of silica is mostly independent of temperature except for the spectral region between 8 to 9 µm, where the actual (physical) reflectance is affected by the temperature-dependent vibrational resonances of $SiO_2$ [20].

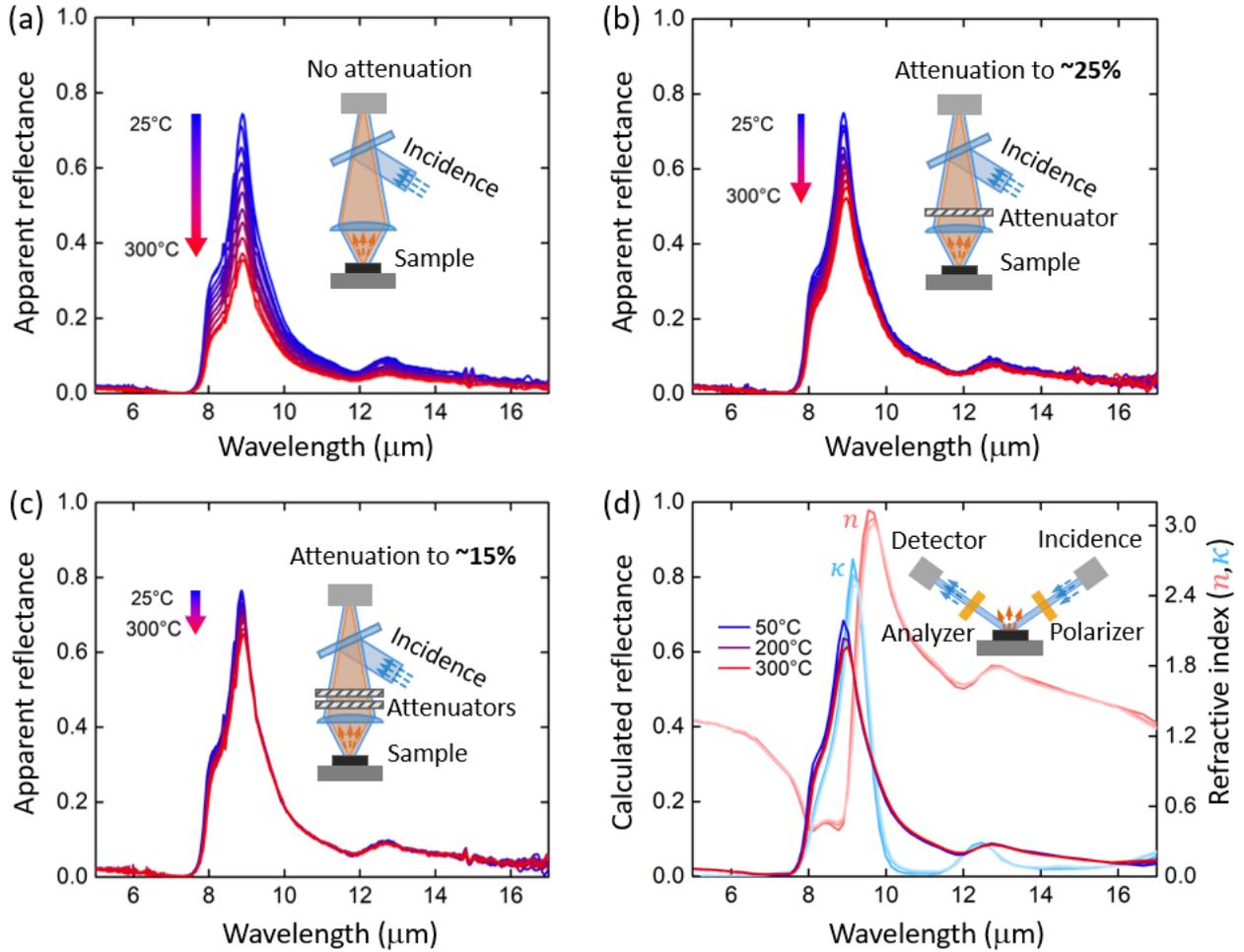

**FIG. 3. Convergence testing using increasing attenuation. (a-c)** Measured apparent reflectance of a fused-silica wafer from 25 to 300 °C using our FTS with (a) no attenuation, (b) attenuation to about 25%, and (c) attenuation to about 15%, respectively. **(d)** Temperature-dependent real ($n$) and imaginary ($\kappa$) parts of the complex refractive index of fused silica extracted from ellipsometry measurements, and the calculated reflectance based on the ellipsometry data. The inset shows the schematic of the ellipsometer used in our measurements with incident angles of 60 and 70°.

For comparison, we also measured and fitted the complex refractive index ($n$ and $\kappa$) of the same fused-silica sample at 50, 200 and 300 °C using infrared variable-angle spectroscopic ellipsometry (IR-VASE) and then calculated its reflectance using Fresnel coefficients, as shown in **Fig. 3(d)**. Note that the ellipsometry measurements were much-less affected by temperature compared to our microscope-based reflectance measurements, because the beam in the ellipsometer was well collimated, and therefore most of the thermal emission (emitted into a large solid angle) could be filtered out by a small aperture in front of the detector which is placed about 25 centimeters away from the sample. Note also that a large solid angle in a microscope measurement is necessary to isolate thermal emission from a small sample, and thus the approach we took with ellipsometry is only applicable to large sample areas. The calculated reflectance is consistent with the FTS reflectance measurements at the highest level of attenuation [i.e., **Fig. 3(c)**].

We also explored a second method for extracting the true temperature-dependent reflectance from the apparent reflectance based entirely on data processing (**Method 2**). This method can be inferred from **Eq. 1**: we can correct the measured interferogram by inferring the mirror-position-independent distortion factor $d(T)$ from experimental data. To demonstrate this approach, we used both the fused-silica wafer we previously used in **Figs. 1** and **3**, and also a polished sapphire wafer (also about 0.5 millimeter thick). **Figure 4(a, d)** shows the apparent reflectance of these samples measured with no attenuation (i.e., the data in **Fig. 4(a)** is the same as **Fig. 1(b)** and **Fig. 3(a)**). In **Fig. 4(b, e)**, we show the ratio $r(\lambda, T)$ of the reflectance measured at temperature $T$ to the reflectance measured at $T_{room} = 25$ °C:

$$r(\lambda, T) = \frac{R(\lambda, T)}{R(\lambda, T_{room})}. \qquad (2)$$

For both samples, $r(\lambda, T)$ has several features. There exist wavelength regions where $r(\lambda, T)$ is temperature-dependent, such as 15 – 16 µm in **Fig. 4(e)**; this temperature dependence is expected to be physical. Other regions are noisy, such as 5 – 8 µm and 15 – 17 µm in **Fig. 4(b)**, resulting from a division by zero where $R(\lambda, T_{room})$ is very small. However, there are also regions within the mid-infrared atmospheric-transparency window where the ratios are almost flat, suggesting that the decrease in $R(\lambda, T)$ is wavelength-independent, and therefore likely nonphysical. We interpret this wavelength-independent region of $r(\lambda, T)$ as representative of the distortion factor $d(T)$ in **Eq. 1**. Note that the "flat ratio" in **Fig. 4(b)** includes a distortion around 12 µm due to the presence of a secondary vibrational resonance; nevertheless, the distortion is small, so we fitted $d(T)$ to the entire 9 – 14 µm region. Thus, we fit these "flat ratios" into constants based on the least-squares method [27], and then the true interferogram can then be calculated by:

$$I_t(x) = \frac{I_m(x,T)}{d(T)}, \quad (3)$$

where $I_t(x)$ and $I_m(x,T)$ represent the true and measured interferogram, respectively. For example, the fitted $d(300\ °C)$ for silica and samphire is 0.49 and 0.57, respectively. Based on **Eq. 3**, we corrected the apparent reflectance of the fused-silica and sapphire wafers, as shown in **Fig. 4(c, f)**. The corrected spectra are mostly temperature-independent, except for the spectral regions corresponding to vibrational resonances. The insets in **Fig. 4(c, f)** show the vibrational-resonance regions of silica and sapphire, respectively, where their optical properties are expected to be temperature-dependent. These characteristic peaks gradually change in value and shift in wavelength with increasing temperatures.

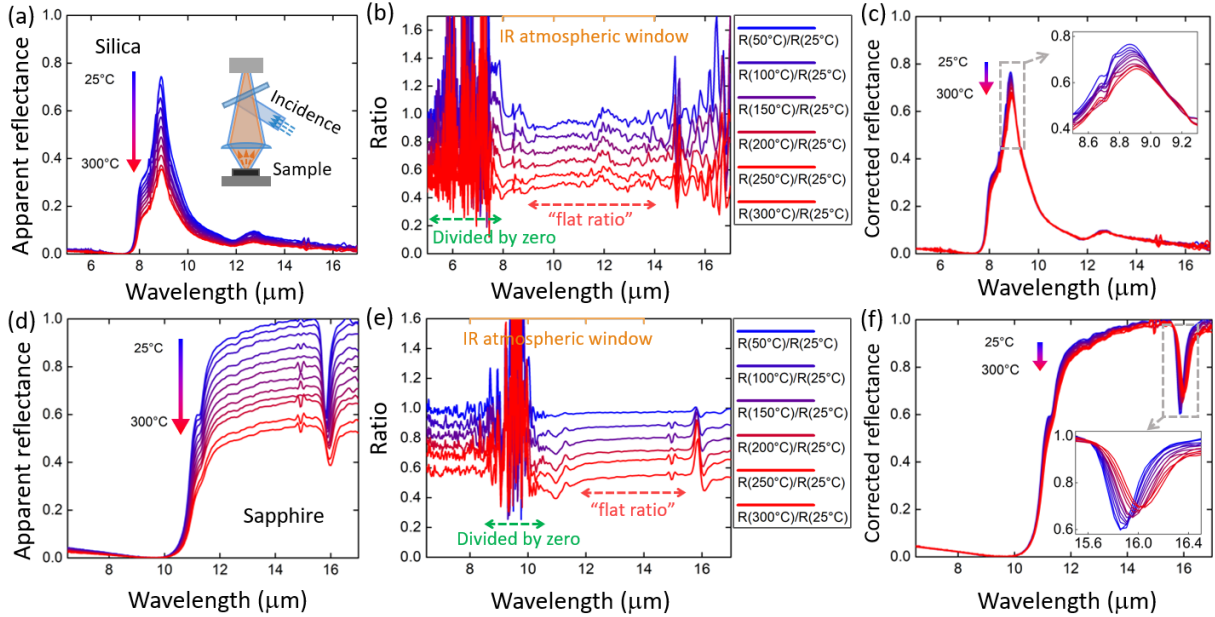

**FIG. 4. Correction of the apparent reflectance assuming the existence of temperature-independent wavelength ranges. (a,d)** Measured apparent reflectance of flat (a) fused-silica and (d) sapphire wafers from 25 to 300 °C using an FTS with no attenuation. **(b,e)** Ratios of spectra measured at different temperatures, i.e.., the ratio between the spectrum at high temperature and the spectrum at 25 °C, for (b) fused silica and (e) sapphire. There exist spectral regions where the ratios are approximately wavelength-independent, and other regions where the ratio is noisy because the reflectance values are very low (resulting in divide-by-zero errors). **(c,f)** Corrected reflectance of (c) fused-silica and (f) sapphire wafers from 25 to 300 °C,

calculated by multiplying the apparent reflectance spectra in (a,d) by the fitted constant ratios in (b,e). The insets show spectral regions near vibrational resonances where the temperature-dependent changes in reflectance are expected to be physical.

In **Fig. 5**, we plotted the reflectance spectra of both fused silica and sapphire corrected using **Method 1** (solid lines) and **Method 2** (dashed lines) at 50 and 300 °C, respectively. The reflectance of silica obtained via ellipsometry measurements and the indirectly calculated reflectance of sapphire via the Kirchhoff's law (i.e., $R(\lambda) = 1 - \varepsilon(\lambda)$) based on our previous direct emissivity measurements [28] are also plotted (dotted lines) for comparison; note that below 7 µm, the sapphire sample becomes partially transparent, and therefore this simple expression for Kirchhoff's law does not apply, resulting in a minor discrepancy. The spectra corrected using **Method 1** and **Method 2** are very consistent for both samples, as evaluated using mean-absolute-percent discrepancies (defined as $\Delta = \frac{1}{n}\sum_{i=1}^{n} \left|\frac{A_i - B_i}{A_i}\right|$). For silica, $\Delta = 1.4\%$ at 50 °C and 2.6% at 300 °C, and for sapphire $\Delta = 1.6\%$ at 50 °C and 3.1% at 300 °C. We note that different references are used for the reflectance and direct-emission measurements (gold for reflectance, carbon-nanotube forest for emission [28]), so imperfect knowledge of the reflectance and emissivity of these references may lead to minor differences in the extracted reflectance spectra. The reflectance of silica based on the ellipsometry measurements is also close to the corrected spectra, though there are some minor differences ($\Delta = 7.6\%$ and 8.7% for silica at 50 and 300°C, respectively) that we attribute to errors in the fitting of $n$ and $\kappa$ and also to the difference in numerical aperture between the ellipsometry and FTS measurements. Note that we did not measure the $n$ and $\kappa$ of the sapphire sample using ellipsometry because sapphire is uniaxial [29], somewhat complicating the fitting process.

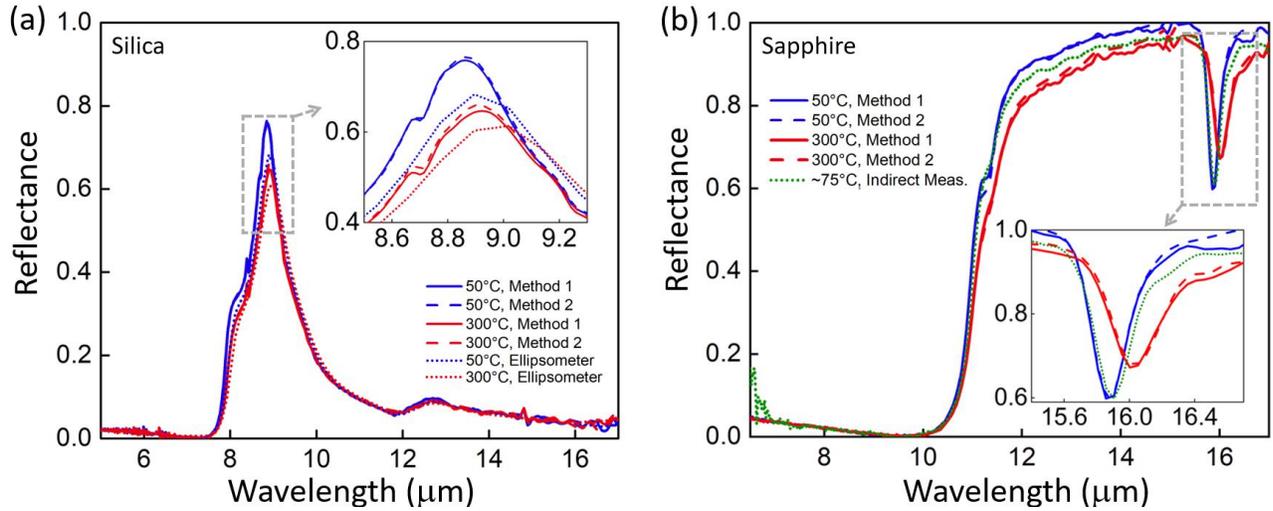

**FIG. 5. Verification of the two correction methods. (a)** Solid lines, dashed lines and dotted lines represent the reflectance of fused silica at 50 and 300 °C obtained based on method 1, method 2, and ellipsometry measurements, respectively. **(b)** Solid lines and dashed lines represent the reflectance of sapphire at 50 and 300 °C obtained based on method 1 and method 2, respectively. The dotted green line represents the indirectly calculated reflectance of sapphire at around 75 °C via Kirchhoff's law based on our previous direct emissivity measurements [28]. The insets show the spectral regions where the temperature-dependent changes in reflectance are expected due to the vibrational resonances.

**Conclusion**

In summary, we observed a nonphysical reduction in the reflectance of samples measured using a Fourier-transform spectrometer (FTS) and infrared microscope for increasing sample temperature. We found that this reduction, which is expected in both reflection and transmission measurements, results from partial detector saturation due to significant thermal emission from the heated samples. The effect is expected to be particularly large for measurement using a large numerical aperture, such as in microscope-based experiments that can be used for small samples, because of the collection of thermal emission over a large solid angle. We proposed and verified two methods to measure or recover the true reflectance (or transmittance) spectra even when the detector saturation is substantial. **Method 1** is a "convergence testing" approach using several levels of attenuation in the setup to reduce detector saturation until it is negligible. The attenuators are placed after the sample to suppress the thermal emission from the sample, though the attenuation can also compromise the signal-to-noise ratio of the measurement, which if needed can be boosted using a lock-in amplifier [30]. **Method 2** applies correction factors by looking at reflectance/transmittance data within spectral regions where the sample has minimal temperature dependence. For materials with weak temperature dependence in at least some spectral regions, this approach can be applied. However, **Method 2** cannot be used for samples with strong temperature-dependent characteristics where no "flat region" over a large wavelength range exists. In practice, the two correction methods can be used independently or in combination, where the latter means implementing sufficient attenuation during the measurements and then fitting the correction factors (expected to be close to 1) to verify or correct the measured spectra.

## Acknowledgements

This research is being performed using funding received from the Department of Energy (DOE) Office of Nuclear Energy's Nuclear Energy University Programs, project 17-13232, grant number DE-NE0008680, and partially by the National Science Foundation (NSF), grant number ECCS-1750341.